\documentclass{article}
\usepackage{amssymb,graphicx,amsmath}
\newtheorem{theorem}{Theorem}
\newtheorem{definition}{Definition}
\newtheorem{lemma}{Lemma}
\def\eop{\hfill$\Box$}
\def\proof{{\em Proof: }}

\begin{document}

\title{Perturbation of coupling matrices and its effect on the
synchronizability in arrays of coupled chaotic systems.}

\author{Chai Wah Wu\\
IBM Research Division,
Thomas J. Watson Research Center\\
P. O. Box 218,
Yorktown Heights, NY 10598, U. S. A.
\thanks{e-mail: chaiwah@watson.ibm.com}}
\maketitle

\vspace{0.1in}

\begin{abstract}
In a recent paper, wavelet analysis was used to perturb the coupling
matrix in an array of identical chaotic systems in order to improve
its synchronization.  As the synchronization criterion is determined
by the second smallest eigenvalue $\lambda_2$ of the coupling matrix,
the problem is equivalent to studying how $\lambda_2$ of the coupling
matrix changes with perturbation.  In the aforementioned paper, a
small percentage of the wavelet coefficients are modified.  However,
this result in a perturbed matrix where every element is modified and
nonzero.  The purpose of this paper is to present some results on the
change of $\lambda_2$ due to perturbation.  In particular, we show
that as the number of systems $n \rightarrow \infty$, perturbations
which only add local coupling will not change $\lambda_2$.  On the
other hand, we show that there exists perturbations which affect an
arbitrarily small percentage of matrix elements, each of which is
changed by an arbitrarily small amount and yet can make $\lambda_2$
arbitrarily large.  These results give conditions on what the
perturbation should be in order to improve the synchronizability in an
array of coupled chaotic systems.  This analysis allows us to prove
and explain some of the synchronization phenomena observed in a
recently studied network where random coupling are added to a locally
connected array.  Finally we classify various classes of coupling
matrices such as small world networks and scale free networks
according to their synchronizability in the limit.
\end{abstract}

\section{Introduction\label{sec:introduction}}
The object of interest in this paper is a coupled array of $n$ identical
systems with linear coupling:
\begin{equation} \dot{x} = \left(\begin{array}{c} f(x_1,t)\\ \vdots \\ f(x_n,t)\end{array}
\right) - (G\otimes D) x
\label{eqn:main}
\end{equation}
where $x = (x_1,\dots,x_n)$.  The coupling matrix $G$ describes the
coupling topology of the array, i.e. $G_{ij}\neq 0$ if there is a
coupling between system $x_i$ and system $x_j$.  If $G_{ij} < 0$ or
$G_{ij} >0$, we call such a coupling element {\em cooperative} or {\em
competitive} respectively.  In this paper we assume that $G$ is a
symmetric zero row sums matrix whose off-diagonal elements are all
nonpositive.  This implies that all coupling is cooperative and all eigenvalues
of $G$ are real. The
matrix $D$ describes the coupling between a pair of systems. In
\cite{wu:chua_synch_arrays_95,pecora_prl_synch_array_1998,wu:synch_book:2002}
it was shown that for a suitable $D$ (such as $D = I$), the
synchronization criterion depends solely on the second smallest
eigenvalue of $G$, which we denote as $\lambda_2(G)$.  In particular,
the larger $\lambda_2(G)$ is, the easier it is to synchronize the
array.  Furthermore, when the single uncoupled system $\dot{y} =
f(y,t)$ is a chaotic system, $\lambda_2$ must necessarily be positive
in order for the array to synchronize.  Therefore we will equate
$\lambda_2$ with synchronizability in the remainder of this paper.  In
\cite{wu:chua_synch_arrays_95} several coupling topologies were
considered.  In particular, for nearest neighbor coupling, $\lambda_2
\rightarrow 0$ as $n\rightarrow\infty$ and thus such an array of
chaotic systems will not synchronize for large $n$.

In \cite{wei-synch-2002} it was shown using wavelet analysis how a
perturbed coupling matrix $\tilde{G} = G + \Delta G$ can be
constructed such that $\lambda_2(\tilde{G})$ is much larger than
$\lambda_2(G)$ and thus the synchronizability is greatly improved.
The perturbation $\Delta G$ can be thought of as the additional
coupling added to the coupling array to affect its synchronization
properties. In \cite{wei-synch-2002}, $\tilde{G}$ is obtained by
modifying a small fraction (the $LL_1$ subspace) of the wavelet
coefficients of $G$.  However, all of the matrix elements of $G$ are
changed in the process, i.e. $\Delta G$ is not sparse.  In particular,
there are no zero elements in $\tilde{G}$, the perturbed array is
fully connected and every chaotic system is connected to every other
chaotic system.  
To illustrate the increase
of $\lambda_2$, in \cite{wei-synch-2002} the $LL_1$ coefficients are
multiplied by a factor of $1101$. This results in a full matrix
$\tilde{G}$ whose off-diagonal elements are up to 8 times larger than
the original matrix $G$. With such drastic changes to $G$, it is not
surprising that $\lambda_2$ will also change dramatically.

Note that Eq. (\ref{eqn:main}) has a minus sign in front of the
term $(G\otimes D) x$ as opposed to a plus sign
in \cite{wei-synch-2002}.  This means that the coupling matrix here is
the negative of the coupling matrix in \cite{wei-synch-2002}.  In
other words, whereas \cite{wei-synch-2002} talks about the second
largest eigenvalue, we consider the second smallest eigenvalue.
Writing the state equation as in Eq. (\ref{eqn:main}) has the benefit
that $\lambda_2$ is positive (rather than negative as in
\cite{wei-synch-2002}).  Furthermore, in this way, 
$G$ and $\lambda_2$ correspond
to the graph-theoretical notions of Laplacian matrix and algebraic
connectivity respectively \cite{wu:chua_graph_theory_synch_95}.  This
is notationally less cumbersome later on when we use
results from the theory of random graphs.

The purpose of this paper is to attempt to answer the question of what
type of matrices $\Delta G$ will make $\lambda_2(\tilde{G})$ much
larger than $\lambda_2(G)$, i.e.  greatly improve the
synchronizability while keeping $\Delta G$ ``small''.  This notion of
smallness could be the sparseness of $\Delta G$ as this refers to the
number of additional coupling added to the original array.  It could
also be the size of the off-diagonal elements of $\Delta G$ as they
relate to the coupling strengths between the chaotic systems in the
array.  We will show that there exists sparse matrices $\Delta G$ with
small elements that will change $\lambda_2(G)$ significantly.  We also
show that uniform coupling will increase the synchronizability the
most.  On the other hand, we show that local coupling,
i.e. $\tilde{G}$ corresponding to a graph with only local connections,
will not synchronize when the number of systems is large.

We also show how this analysis allows us to prove and explain the
change in synchronizability observed in a coupled array where random
coupling elements are added to a nearest neighbor coupled array.

Finally we will classify several classes of coupling matrices
considered in the literature according to a synchronizability measure,
and show that the more random and less ordered the coupling graph is,
the higher the synchronizability.

\section{Super-additivity of $\lambda_2$}
\begin{definition}
$W$ is the class of symmetric matrices that have zero row sums and
nonpositive off-diagonal elements.
\end{definition}
It can be shown that matrices in $W$ are positive semi-definite and
that $0$ is a simple eigenvalue if and only if the matrix is irreducible
\cite{wu:chua_synch_arrays_95}.  We assume that the coupling matrices 
$G$ and $\tilde{G}$ are in $W$.

\begin{lemma} \label{lem:lambda}
If $A$ and $B$ are in $W$, then $\lambda_2(A+B) \geq \lambda_2(A) +
\lambda_2(B)$.
\end{lemma}
\proof see \cite{wu:chua_synch_arrays_95}.\eop

This lemma shows that $\lambda_2(\tilde{G}) \geq \lambda_2(G) +
\lambda_2(\Delta G)$ when $\Delta G$ is also a matrix in $W$.
Therefore if we can find $\Delta G$ such that $\lambda_2(\Delta G)$ is
large, then we can increase $\lambda_2(\tilde{G})$ and thus the
synchronizability.

A corollary to Lemma \ref{lem:lambda} is that if $A$, $B$ are matrices
in $W$ and every off-diagonal element of $A$ is smaller in absolute
value than $B$, then $\lambda_2(A)\leq\lambda_2(B)$.  This means that
adding additional cooperative coupling elements ($G_{ij} < 0$) can
only increase $\lambda_2$ and improve the synchronizability.

Let us consider three reasonable constraints on the possible coupling
matrix $G$ that we can choose:
\begin{enumerate}
\item $G \in W$.
\item the number of nonzero coupling elements is a percentage of the total number of coupling elements, i.e. only a percentage of the
off-diagonal elements of $G$ are nonzero.
\item the absolute value of
each off-diagonal element is less than some fixed number $B >0$.
\end{enumerate}
Under these constraints, the above discussion shows that the matrix
$G$ with the largest $\lambda_2$ must necessarily be matrices in $W$
where the off-diagonal elements are either $0$ or $-B$.  We call such
a coupling configuration a {\em uniform} coupling as the coupling
between any pair of systems is either zero or the same\footnote{This
notion of uniform coupling is different than the notion used by some
authors to denote fully connected coupling.}.  Thus given a bound on
the size of each coupling element, the uniform coupling is the best
coupling configuration in terms of synchronizability.

This motivates us to study the class of matrices in $W$ where the
off-diagonal elements is either 0 or -1.  Such a matrix is the
Laplacian matrix of the underlying graph: vertex $i$ is connected to
vertex $j$ if and only if $G_{ij} \neq 0$ and $\lambda_2$ corresponds
to the algebraic connectivity of the graph \cite{fiedler:algebraic_connectivity_73}.  This
allows us to use results in graph theory to study
$\lambda_2(\tilde{G})$.  In particular, in the rest of the paper, we
will mainly consider coupling matrices which is a positive multiple of the
Laplacian matrix of the underlying graph and the goal is to find
appropriate coupling graphs whose algebraic connectivity are large.

\section{Local coupling} \label{sec:local}
Consider the following model of local coupling
\cite{wu:observer_synch:2001}: let the vertices of the coupling graph
be the points in the hypercube $[0,r]^d$ with integer coordinates,
i.e. the vertices are the lattice points inside the hypercube
$[0,r]^d$ generated by the standard basis vectors.  Let the systems in the
array be located on these vertices.  We say this array is {\em locally
connected} if there exists a fixed $\delta$ such that all the systems
which are connected to each system $x$ lies in the
$\delta$-neighborhood of $x$.  This notion of local coupling includes
examples such as nearest neighbor coupling and grid graphs, but also
more general local coupling configurations where the vertex degree is
bounded but not necessarily constant.

In \cite{wu:observer_synch:2001} it was shown that:
\begin{lemma} \label{lem:local}
For fixed $d$ and
$\delta$, a locally connected array will have $\lambda_2 \rightarrow 0$
as $n\rightarrow \infty$ (i.e. $r\rightarrow \infty$).  This is true
even if the hypercube is considered as a hypertorus. 
\end{lemma}

This means that locally connected arrays will have low
synchronizability for large $n$.  In particular, locally connected
arrays of chaotic systems will not synchronize for large $n$.  This
means that $\Delta G$ should have nonlocal coupling in order to
improve the synchronizability of the array. Next we show that random
coupling is some sense the best candidate for such nonlocal coupling.

\section{Random coupling} \label{sec:random}

How much coupling do we need to add in order to make $\lambda_2$ large
while keeping the size of the off-diagonal element small?  In
\cite{fiedler:algebraic_connectivity_73} it was shown that:

\begin{lemma}\label{lem:degree}
For a non-complete graph $\lambda_2$ is less than or equal to the
minimum vertex degree\footnote{For a complete graph, $\lambda_2(K_n) =
n$.}.
\end{lemma}

In particular, for a $k$-regular (non-complete) graph, $\lambda_2 \leq
k$.  Therefore, if we want $\lambda_2$ to be large we need each vertex
to connect to a large number of other vertices, i.e. $\Delta G$ cannot
be too sparse.  Regular random graphs appear to be asymptotically
optimal in this respect by Lemma \ref{lem:degree} and the following
lemma.

\begin{lemma}\label{lem:random}
There exists a constant $0 < c < 1$ such that almost every $k$-regular random
graph satisfies $\lambda_2 > ck$ as $n\rightarrow \infty$ for large enough $k$.
For even $k$, $c$ can be chosen close to $1$.
\end{lemma}
\proof See \cite{wu:observer_synch:2001}.\eop

\begin{figure}
\centerline{\includegraphics[width=4.4in]{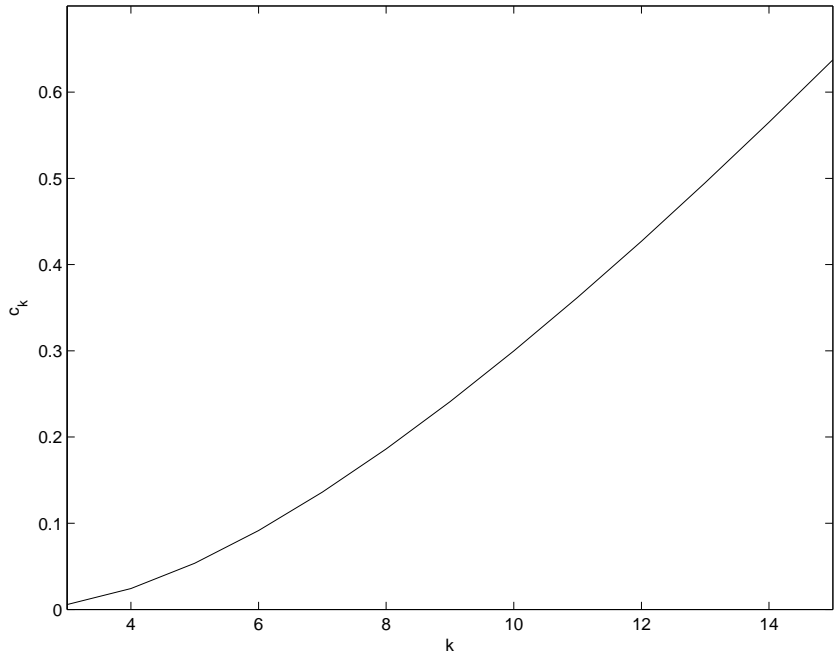}}
\caption{Plot of lower bound $c_k$ of the second smallest eigenvalue of
$k$-regular random graphs for $3\leq k\leq 15$.
}
\label{fig:lower}
\end{figure}

For small $k$, it can be
shown that for $k\geq 3$, there is a constant $c_k > 0$ such that
almost every $k$-regular random graph satisfies $\lambda_2 \geq c_k$
as $n\rightarrow \infty$.  In particular, since $\lambda_2 \geq k -
\sqrt{k^2-i^2}$
\cite{mohar:isoperimetric_graphs_1989,wu:chua_graph_theory_synch_95}
where $i$ is the isoperimetric number of the graph, and using the
results in \cite{bollobas:isoperimetric_random:1988},
a graph of $c_k$ for $k$ between $3$ and $15$ is shown in Figure
\ref{fig:lower}.

This allows us to prove the following theorem which says that there is
an $n$ by $n$ coupling matrix $\Delta G$ with an arbitrarily small
percentage of nonzero off-diagonal elements, each of which is
arbitrarily small, and which has arbitrarily large $\lambda_2$ when
$n$ is large enough.

\begin{theorem}
Given $0<p\leq 1$, $r>0$, and $s$, for all large enough $n$, 
there exists a matrix $\Delta G$ in $W$ such that
\begin{itemize}
\item $\lambda_2 > s$,
\item the fraction of nonzero elements of $\Delta G$ is $p$,
\item each off-diagonal element of $\Delta G$ is less
than $r$.
\end{itemize}
\end{theorem}
\proof Choose $k$ an integer larger than $\frac{s}{cr}$. By Lemma
\ref{lem:random} for large enough $n$ and $k$, almost every random $k$-regular
graph will have $\lambda_2 > \frac{s}{r}$.  Choose large $n$ such that
$n > \frac{k}{p}$.  Let $A$ be the Laplacian matrix of such a graph.
Then $\Delta G = rA$ will suffice. \eop

So far, the random graph corresponding to $\Delta G$ is required to be
a regular graph.  Next we show that this requirement can be relaxed.
In \cite{wang:synch_small_world_2002} a network model is presented
where a nearest neighbor configuration is modified by adding a
fraction $0\leq p\leq 1$ of random edges out of all possible
edges\footnote{The authors of \cite{wang:synch_small_world_2002}
erroneously associated this model with the small world model in
\cite{newman-small_world-1999}. In \cite{newman-small_world-1999} the
number of added edges is a fraction $p$ of the existing edges in the
nearest neighbor coupling.  Therefore the number of edges in
\cite{newman-small_world-1999} is much less.  In particular, for
$p=1$, the graph in \cite{wang:synch_small_world_2002} is fully
connected, whereas in \cite{newman-small_world-1999} the number of
edges is only twice the number of edges of the nearest coupling graph.
In particular for the one-dimensional nearest neighbor coupling, by
Lemma \ref{lem:degree} the algebraic connectivity of the model in
\cite{newman-small_world-1999} is at most 2, whereas it is unbounded
in \cite{wang:synch_small_world_2002} as $n\rightarrow\infty$.}.  In
\cite{wang:synch_small_world_2002} it was shown via numerical
experiments that $\lambda_2(\tilde{G})$ increases in a near linear
fashion with respect to $p$ and to $n$.  We prove that this is true in
general.  Let $G$ be a locally connected graph and $\Delta G$ is
constructed as in \cite{wang:synch_small_world_2002} by picking
randomly a fraction $p$ of all possible edges.

First note that by Lemmas \ref{lem:lambda} and \ref{lem:degree},
$\lambda_2 (\tilde{G}) \leq pn + q$ where $q$ is the average number of
connections of $G$.  In \cite{juhasz:random_alg_conn:1991} it was
shown that $\lambda_2(\Delta G)$ is equal to $pn +
o(n^{\frac{1}{2}+\epsilon})$ in probability for any $\epsilon >0$.
Combining this with Lemmas \ref{lem:lambda} and \ref{lem:local}, it's
clear that $\lambda_2 (\tilde{G})$ increases as $o(pn)$ in
probability as $n\rightarrow\infty$, i.e. linear in $p$ and in $n$.

Thus in general, given a locally connected array, the addition of a
fraction of random edges out of all possible edges will increase the
synchronizability by an arbitrarily large amount as the number of
vertices increases.  Note that the synchronizability of the
random edges dominates that of the locally connected graph. 

\subsection{Example}
To compare the results in this section with the results in
\cite{wei-synch-2002}, consider the example in \cite{wei-synch-2002}
that was discussed in Section \ref{sec:introduction} where an array of
64 oscillators with nearest neighbour coupling is perturbed by
multiplying the $LL_1$ wavelet coefficients of the coupling matrix $G$
by a factor of $1101$.  This increases $\lambda_2$ from $0.0096$ to
$0.1522$.  However, all entries of the 64 by 64 matrix $G$ are
changed, with the off-diagonal entries of the perturbation matrix
$\Delta G$ ranging from $-7.6$ to $9.7$.  On the other hand, suppose
we want to change only $5$ percent of the off-diagonal entries of $G$.
By choosing a random graph with $100$ edges and corresponding
Laplacian matrix $L$, we construct a perturbation matrix $\Delta G =
0.0665L$ which also increases $\lambda_2$ from $0.0096$ to $0.1522$.
However, in this case, only $5$ percent of the off-diagonal entries of
$G$ are perturbed, and the off-diagonal entries of the perturbation
matrix $\Delta G$ is either $-0.0665$ or $0$.  Thus the improvement in
synchronization is the same as in \cite{wei-synch-2002}, but the
perturbation matrix is much smaller, both in the sense of the number
of nonzero elements and in the sense of the magnitude of the nonzero
elements.

\section{Comparison of various classes of coupling graphs}
In this section we compare several classes of coupling graphs that
have been proposed in the literature for modeling complex networks,
both man-made (the internet or social networks) and occuring in nature
(neural networks).  We can categorize their synchronizability by
studying the synchronizability measure $s(m) = \lim_{n\rightarrow\infty}
\frac{\lambda_2}{m}$ when it exists, where $m$ is the average number
of connections defined as 2 times the number of edges divided by the number of
vertices, i.e. $m$ is the number of nonzero off-diagonal elements of the Laplacian matrix divided by $n$.  For $r$-regular graphs, $m = r$.  
By Lemma \ref{lem:degree}, $0\leq s(m)\leq 1$.
We will only consider the cases $m\geq 3$. 

\subsection{Small world networks}
In \cite{newman-small_world-1999} a small-world network model is
introduced where starting from a locally connected graph with $m$
edges, $pm$ edges ($0 <p\leq 1$) are added randomly.  Without the
random edges $\lambda_2\rightarrow 0$ as $n\rightarrow\infty$ by Lemma
\ref{lem:local}.  In the cases considered in
\cite{newman-small_world-1999}, $m$ grows linearly with $n$, i.e. $m =
kn$.  If we modified this model such that the added random edges form
a regular random graph\footnote{This requires that $2kp$ is an
integer.}, then the discussion in Section \ref{sec:random} shows that
$\lambda_2$ is bounded away from zero as $n\rightarrow\infty$ when $kp
\geq \frac{3}{2}$.  In particular, Figure \ref{fig:lower} shows that
$s(m)\geq 0.0018$.  Thus by adding the random edges, there is a
coupling strength factor $\epsilon > 0$ such that when $D$ is replaced
by $\epsilon D$, the coupled array will synchronize independent of
$n$.

\subsection{Scale-free random networks}
\begin{figure}
\centerline{\includegraphics[width=4.4in]{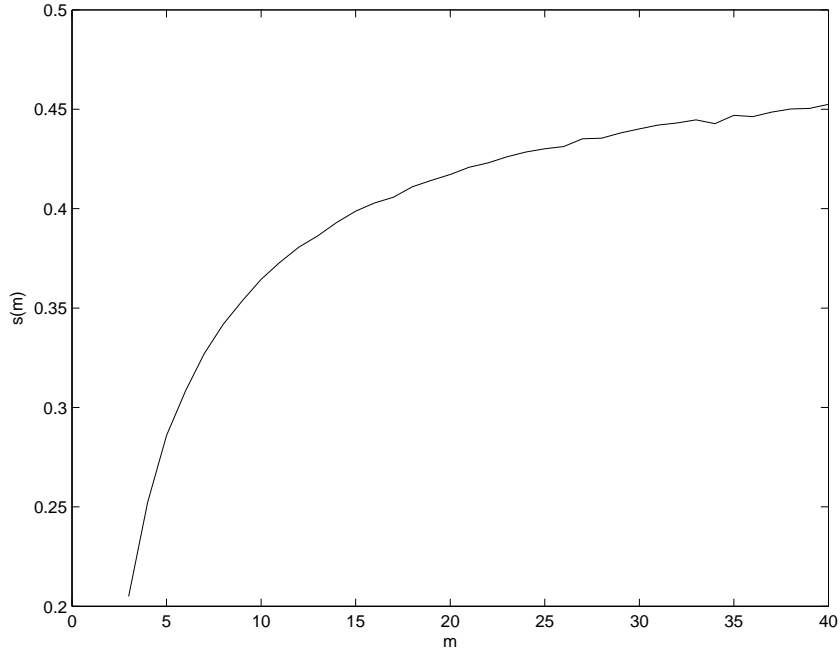}}
\caption{Plot of the synchronizability measure $s(m)$ for scale free
random networks.  We see that $s(m)$ is increasing and approaches
close to the upper bound $\frac{1}{2}$.  }
\label{fig:scalefreesm}
\end{figure}

In \cite{albert_scale-free_2000} a scale-free random network model of
the internet is presented where the graph is built up as follows:
Starting from a initial set of $k_0$ (usually set at $k_0 = k$)
vertices, at each iteration a new vertex is added with $k$ new edges
connecting this vertex to $k$ existing vertices.  Thus the average
connections is $m = 2k$ as $n\rightarrow\infty$.  The probability that
the new vertex is connected to vertex $i$ is proportional to the
vertex degree of vertex $i$. This results in a graph whose vertex
degree distribution follows a power law.  Since there is always at
least one vertex with degree $k$, $\lambda_2 \leq k$ and $s(m)\leq
\frac{1}{2}$ by Lemma \ref{lem:degree}.  Computer
simulations\footnote{We differ slightly from
\cite{albert_scale-free_2000} in our construction of the scale-free
networks.  In \cite{albert_scale-free_2000} the initial configuration
is $k_0$ vertices with no edges.  After many iterations, this could
still result in some of these initial $k_0$ vertices having degree
less than $k$, even though every newly added vertex has degree at
least $k$.  Since even one vertex with degree 1 results in $\lambda_2$
to drop below $1$, we would like to have every vertex of the graph to
have degree at least $k$. Therefore we will start with the initial
$k_0$ vertices fully connected to each other when $k_0 = k$.  For the
case $k_0 > k$, we require that the initial $k_0$ vertices are
connected with each other with vertex degrees at least $k$.  This will
guarantee that the vertex degree of every vertex is at least $k$.}
show that for a fixed $m$, $\lambda_2$ of such graphs appear to
increase monotonically as a function of $n$.  If this is the case,
then since $\lambda_2$ is bounded by $\frac{m}{2}$, this means that
$\lambda_2$ converges to a constant value $\lambda_2^\infty(m)$ as
$n\rightarrow\infty$.  We computed via simulations the value of $s(m)
= \frac{\lambda_2^\infty(m)}{m}$ for various $m$ as shown in Figure
\ref{fig:scalefreesm}.  We see that $s(m)$ is increasing and
converging to close to the upper bound $\frac{1}{2}$.  As the
scale-free network is not as random as a random network, its
synchronizability measure is also less.

For locally connected graphs, $s(m) = 0$ for all $m$ by Lemma
\ref{lem:local}.  For both small-world networks and the model in
\cite{wang:synch_small_world_2002} $s(m) \approx 1$ for large $m$.
The main difference is that in small-world networks, $m$ (and thus
also $\lambda_2$) is small and bounded, whereas in
\cite{wang:synch_small_world_2002}, $m = pn$ and therefore $\lambda_2$
is unbounded.  These results are summarized in Table
\ref{tbl:classes}.

\begin{table}
\begin{center}
\begin{tabular}{|l|l|} \hline
Graph class & Synchronizability measure $s(m)$ \\
\hline\hline
Locally connected graphs \cite{wu:observer_synch:2001} & $0$ \\ \hline
Scale-free networks \cite{albert_scale-free_2000} & $ \approx 0.5$ as $m\rightarrow\infty$ \\ \hline
Small world networks \cite{newman-small_world-1999} & $\geq 0.0018$,\quad $\approx 1$ as $m\rightarrow\infty$ \\ \hline
Model in \cite{wang:synch_small_world_2002} & $s(m)\approx 1$, $\lambda_2\rightarrow\infty$ as $n\rightarrow\infty$ \\ \hline\hline
\end{tabular}
\caption{Several classes of coupling graphs and their synchronizability measure $s(m)$.}\label{tbl:classes}
\end{center}
\end{table}

Thus we can order these classes of matrices as follows in terms of
increasing synchronizability: locally connected graphs, scale-free
networks, small-world networks, and the model in
\cite{wang:synch_small_world_2002}.  We see that the more disorder
and randomness the graph exhibits, the higher the synchronizability.

\section{Conclusions}
We study the conditions under which perturbation of the coupling in a
coupled array results in improved synchronizability by reducing it to
the analysis of the second smallest eigenvalue of the coupling matrix.
We show that local coupling and random coupling form two extremes in
their effect on synchronizability.  In particular, local coupling will
not improve synchronizability in the limit and random coupling can
change the synchronizability using arbitrarily small percentage of
coupling elements whose strengths are arbitrarily small.  As graphs
become less ordered and more random, we see that the synchronizability
increases.  Furthermore, for graphs which is constructed by adding a
random component to a structured component, the super-additivity of
$\lambda_2$ allows us to give a lower bound for the synchronizability
of graphs by adding the synchronizability of each of the components.
In particular, when the structured component is a locally connected
graph, the synchronizability of the random component dominates the
synchronizability of the locally connected component.

\bibliography{chua_ckt2,chaos,secure,synch,misc,stability,cml,algebraic_graph,graph_theory,control,optimization,adaptive,top_conjugacy,ckt_theory,cnn2,matrices,chaos_comm}

\begin{thebibliography}{10}

\bibitem{wu:chua_synch_arrays_95}
C.~W. Wu and L.~O. Chua, ``Synchronization in an array of linearly coupled
  dynamical systems,'' {\em IEEE Transactions on Circuits and Systems--I:
  Fundamental Theory and Applications}, vol.~42, no.~8, pp.~430--447, 1995.

\bibitem{pecora_prl_synch_array_1998}
L.~M. Pecora and T.~L. Carroll, ``Master stability functions for synchronized
  coupled systems,'' {\em Physical Review Letters}, vol.~80, no.~10,
  pp.~2109--2112, 1998.

\bibitem{wu:synch_book:2002}
C.~W. Wu, {\em Synchronization in coupled chaotic circuits and systems}.
\newblock World Scientific, 2002.

\bibitem{wei-synch-2002}
G.~W. Wei, M.~Zhan, and C.-H. Lai, ``Tailoring wavelets for chaos control,''
  {\em Physical Review Letters}, vol.~89, no.~28, p.~284103, 2002.

\bibitem{wu:chua_graph_theory_synch_95}
C.~W. Wu and L.~O. Chua, ``Application of graph theory to the synchronization
  in an array of coupled nonlinear oscillators,'' {\em IEEE Transactions on
  Circuits and Systems--I: Fundamental Theory and Applications}, vol.~42,
  pp.~494--497, Aug. 1995.

\bibitem{fiedler:algebraic_connectivity_73}
M.~Fiedler, ``Algebraic connectivity of graphs,'' {\em Czechoslovak
  Mathematical Journal}, vol.~23, no.~98, pp.~298--305, 1973.

\bibitem{wu:observer_synch:2001}
C.~W. Wu, ``Synchronization in arrays of coupled nonlinear systems: Passivity,
  circle criterion and observer design,'' {\em IEEE Transactions on Circuits
  and Systems--I: Fundamental Theory and Applications}, vol.~48, no.~10,
  pp.~1257--1261, 2001.

\bibitem{mohar:isoperimetric_graphs_1989}
B.~Mohar, ``Isoperimetric numbers of graphs,'' {\em Journal of Combinatorial
  Theory, Series B}, vol.~47, pp.~274--291, 1989.

\bibitem{bollobas:isoperimetric_random:1988}
B.~Bollob\'{a}s, ``The isoperimetric number of random regular graphs,'' {\em
  European Journal of Combinatorics}, vol.~9, pp.~241--244, 1988.

\bibitem{wang:synch_small_world_2002}
X.~F. Wang and G.~Chen, ``Synchronization in small-world dynamical networks,''
  {\em International Journal of Bifurcation and Chaos}, vol.~12, no.~1,
  pp.~187--192, 2002.

\bibitem{newman-small_world-1999}
M.~E.~J. Newman and D.~J. Watts, ``Scaling and percolation in the small-world
  network model,'' {\em Physical Review E}, vol.~60, no.~6, pp.~7332--7342,
  1999.

\bibitem{juhasz:random_alg_conn:1991}
F.~Juh\'{a}sz, ``The asymptotic behaviour of {F}iedler's algebraic connectivity
  for random graphs,'' {\em Discrete Mathematics}, vol.~96, pp.~59--63, 1991.

\bibitem{albert_scale-free_2000}
A.-L. Barab\'{a}si, R.~Albert, and H.~Jeong, ``Scale-free characteristics of
  random networks: the topology of the world wide web,'' {\em Physica A},
  vol.~281, pp.~69--77, 2000.

\end{thebibliography}
\end{document}